\documentclass[3p,times]{elsarticle}
\usepackage{amssymb}
\usepackage{graphicx}
\usepackage{amsmath}
\newcommand{\beq}{\begin{equation}}
\newcommand{\eeq}{\end{equation}}
\newcommand{\beqa}{\begin{eqnarray}}
\newcommand{\eeqa}{\end{eqnarray}}
\newcommand{\bseq}{\begin{subequations}}
\newcommand{\eseq}{\end{subequations}}

\def\x{{\boldsymbol x}}

\def\p{{\boldsymbol p}}
\def\0{{\boldsymbol 0}}

\def\simle{\mathrel{\rlap{\raise 0.511ex \hbox{$<$}}{\lower 0.511ex 
\hbox{$\sim$}}}}
\def\simge{\mathrel{ \rlap{\raise 0.511ex 
\hbox{$>$}}{\lower 0.511ex \hbox{$\sim$}}}}

\begin{document}

\begin{frontmatter}
\title{Heavy quark dynamics in the QGP: $R_{AA}$ and $v_2$ from RHIC to LHC}
\author{W.M. Alberico$^{3,4}$, A. Beraudo$^{1,2,3,4}$, A. De Pace$^4$, A. Molinari$^{3,4}$, M. Monteno$^4$, M. Nardi$^4$ and F. Prino$^4$}
\address{$^1$Centro Studi e Ricerche \emph{Enrico Fermi}, Piazza del Viminale 1, Roma, ITALY}
\address{$^2$Physics Department, Theory Unit, CERN, CH-1211 Gen\`eve 23, Switzerland}
\address{$^3$Dipartimento di Fisica Teorica dell'Universit\`a di Torino, via P.Giuria 1, I-10125 Torino, Italy}
\address{$^4$Istituto Nazionale di Fisica Nucleare, Sezione di Torino, via P.Giuria 1, I-10125 Torino, Italy}

\begin{abstract}
We study the stochastic dynamics of $c$ and $b$ quarks in the hot plasma produced in nucleus-nucleus collisions at RHIC and LHC, providing results for the nuclear modification factor $R_{AA}$ and the elliptic flow coefficient $v_2$ of the single-electron spectra arising from their semi-leptonic decays. The initial $Q\overline{Q}$ pairs are generated using the POWHEG code, implementing pQCD at NLO.
For the propagation in the plasma we develop a relativistic Langevin equation (solved in a medium described by hydrodynamics) whose transport coefficients are evaluated through a first-principle calculation. Finally, at $T_c$, the heavy quarks are made hadronize and decay into electrons: the resulting spectra are then compared with RHIC results. Predictions for LHC are also attempted.
\end{abstract}

\begin{keyword}
Quark Gluon Plasma \sep heavy quarks \sep Langevin equation\sep transport coefficients
\end{keyword}
\end{frontmatter}
\section{Introduction}
Heavy quarks, produced in initial hard processes, allow to perform a ``tomography'' of the medium created in the relativistic heavy-ion collisions at RHIC and (soon) at LHC. The modification of their spectra provides information on the properties (encoded into few transport coefficients) of the matter (hopefully a thermalized Quark Gluon Plasma) crossed before hadronizing and giving rise to experimental signals: so far, the electrons from their semi-leptonic decays. We employ an approach based on the relativistic Langevin equation, assuming that medium-modifications of the initial heavy-quark spectrum arise from the effects of many independent random collisions. As a final outcome we provide results for the $R_{AA}$ and $v_2$ of non-photonic electrons from heavy-flavor decays measured at RHIC and attempt predictions for LHC. Further results can be found in Refs.~\cite{lange_hot,lange_pra}. For similar studies see Refs.~\cite{hira,rapp,tea,aic,das}.
\section{The relativistic Langevin equation}
The usual Langevin equation can be generalized to the relativistic case~\cite{lange}, providing a tool to study the propagation of $c$ and $b$ quarks in the QGP. The variation of the heavy-quark momentum in the time-interval $\Delta t$
\beq\label{eq:lange_r_d}
\frac{\Delta p^i}{\Delta t}=-\eta_D(p)p^i+\xi^i(t),
\eeq
is given by the sum of a \emph{deterministic} friction term and a \emph{stochastic} noise term $\xi^i(t)$, which is completely determined by its two-point temporal correlator  
\beq\label{eq:noise1}
\langle\xi^i(t)\xi^j(t')\rangle=b^{ij}(\p)\delta(t-t'),\quad{\rm with}\quad
b^{ij}(\p)\equiv \kappa_L(p)\hat{p}^i\hat{p}^j+\kappa_T(p)
(\delta^{ij}-\hat{p}^i\hat{p}^j).
\eeq
The latter involves the transport coefficients $\kappa_T(p)\!\equiv\!\frac{1}{2}\frac{\langle \Delta p_T^2\rangle}{\Delta t}$ and $\kappa_L(p)\!\equiv\!\frac{\langle \Delta p_L^2\rangle}{\Delta t}$,
representing the average transverse and longitudinal squared-momentum acquired per unit time by the heavy quark due to the collisions suffered in the medium.
Finally, as in the non-relativistic case, the friction coefficient $\eta_D(p)$ is fixed in order to insure the approach to thermal equilibrium.
In the Ito discretization \cite{ito} of Eq.~(\ref{eq:lange_r_d}) one has:
\beq
\eta_D^{\text{Ito}}(p)=\frac{\kappa_L(p)}{2TE} 
-\frac{1}{E^2}\left[(1-v^2)\frac{\partial\kappa_L(p)}{\partial v^2}+
\frac{d-1}{2}\,\frac{\kappa_L(p)-\kappa_T(p)}{v^2}\right], 
\eeq
Eq.~(\ref{eq:lange_r_d}) has then to be solved in the evolving medium produced in the heavy-ion collisions and described by ideal/viscous hydrodynamics. For this purpose the output of two independent hydro codes~\cite{kolb1,rom1,rom2} are exploited. Details on the employed procedure can be found in Refs.~\cite{lange_hot,lange_pra}. 
\section{Evaluation of the transport coefficients}
\begin{figure}
\begin{center}
\includegraphics[clip,width=0.495\textwidth]{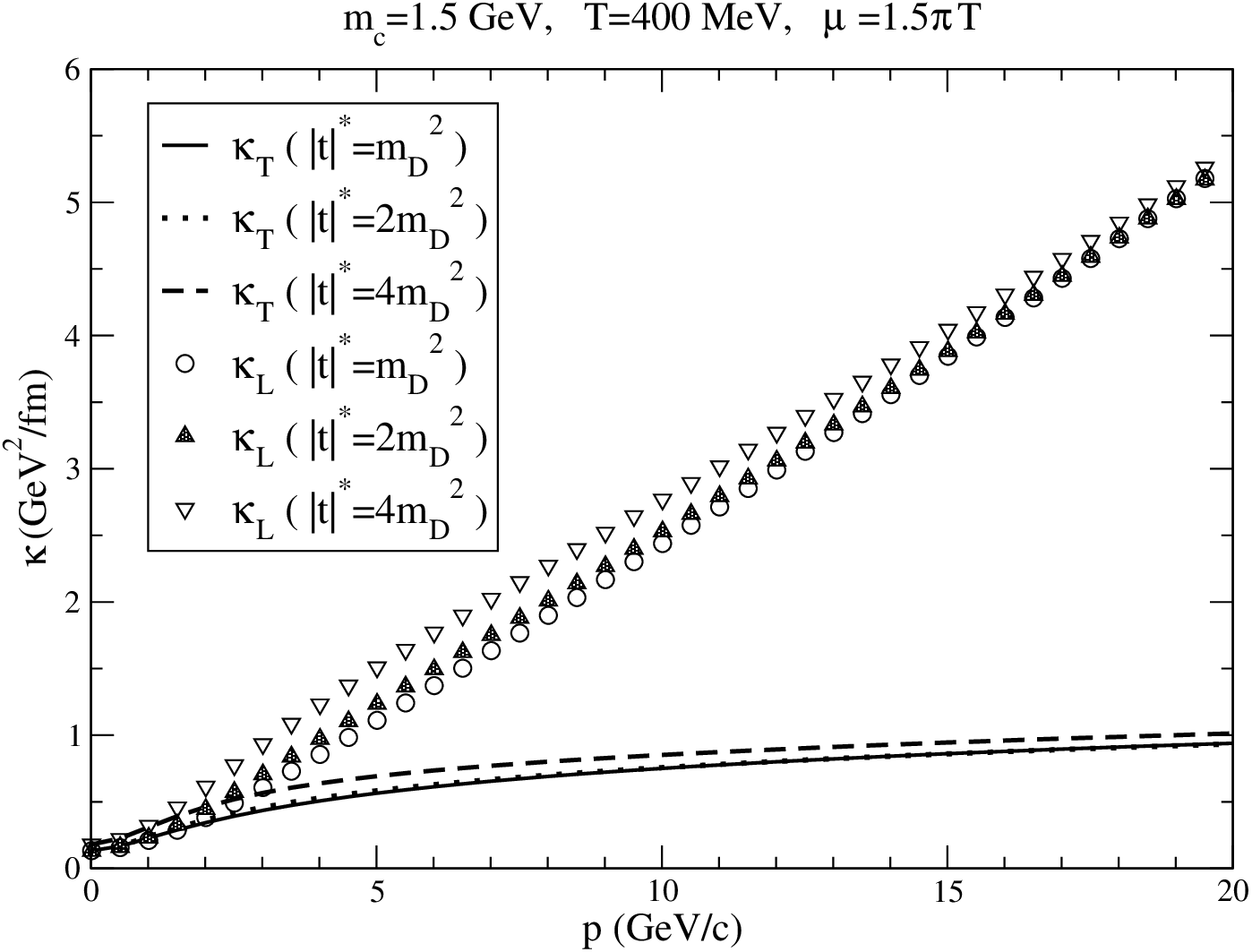}
\includegraphics[clip,width=0.495\textwidth]{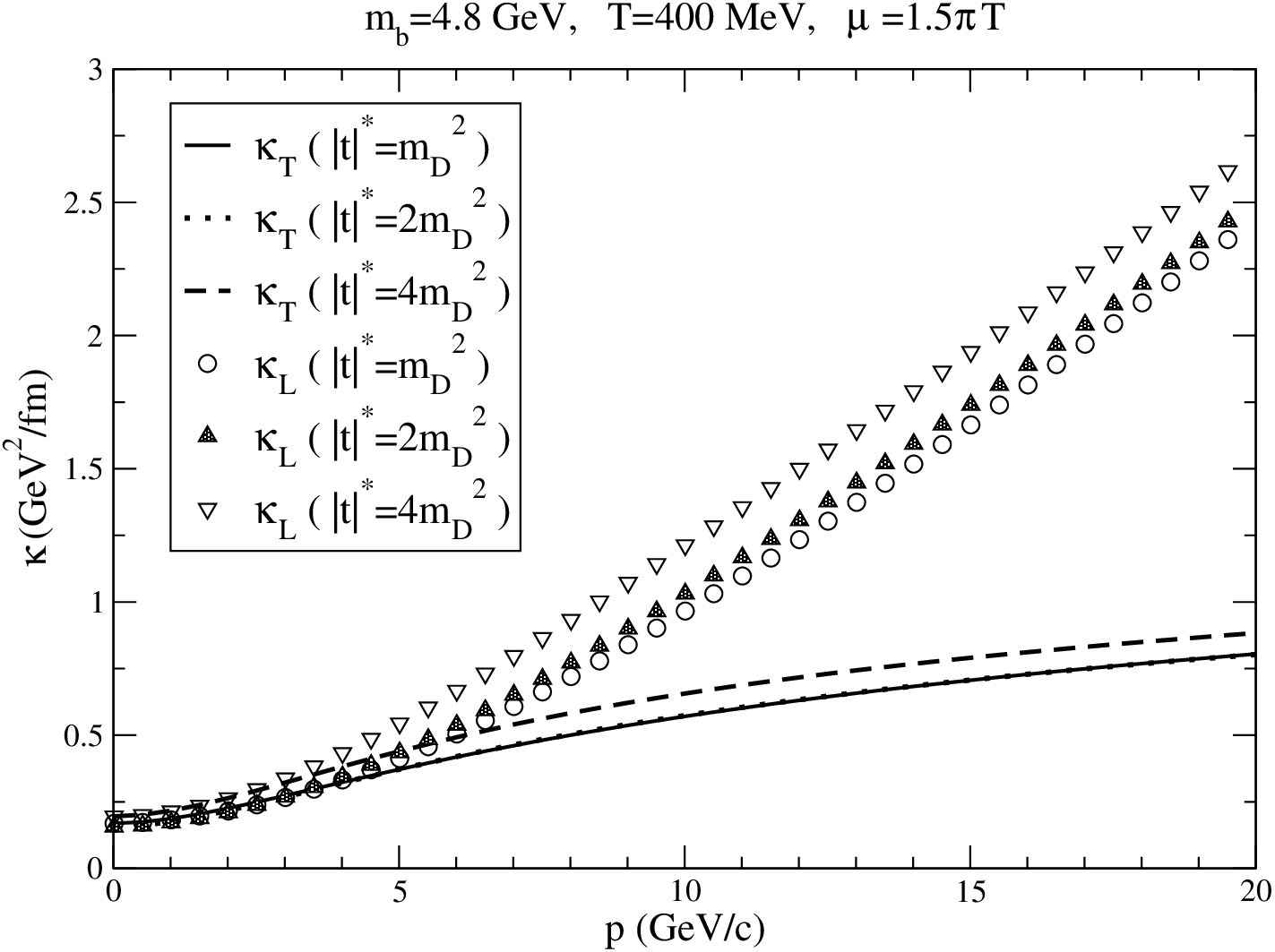}
\caption{The transport coefficients $\kappa_{T/L}$ of $c$ and $b$ quarks after summing the soft and hard contributions. The dependence on the cutoff $|t|^*$ is mild. The coupling $g$ was evaluated at the ``central value'' of our systematic scan $\mu\!=\!1.5\pi T$.} 
\label{fig:transport}
\end{center}
\end{figure}

The transport coefficients $\kappa_{T/L}$ are evaluated according to the procedure presented in Refs.~\cite{lange_hot,lange_pra}. Following Ref.~\cite{pei} we introduce an intermediate cutoff $|t|^*\!\sim\!m_D^2$ ($t\!\equiv\!(P'\!-\!P)^2$) separating hard and soft scatterings. The contribution of hard collisions ($|t|\!>\!|t|^*$) is evaluated through a kinetic pQCD calculation of the processes $Q(P)q_{i/\bar i}\!\to\! Q(P')q_{i/\bar i}$ and $Q(P)g\!\to\! Q(P')g$~\cite{com}. On the other hand in soft collisions ($|t|\!<\!|t|^*$) the exchanged gluon feels the presence of the plasma. A resummation of medium effects is thus required and this is provided by the Hard Thermal Loop approximation. The final result is given by the sum of the two contributions $\kappa_{T/L}(p)=\kappa_{T/L}^{\rm hard}(p)+\kappa_{T/L}^{\rm soft}(p)$ and its explicit expression can be found in Ref.~\cite{lange_hot}. In Fig.~\ref{fig:transport} we display the behavior of the transport coefficients of $c$ and $b$ quarks. The sensitivity to the value of the intermediate cutoff $|t|^*$ is quite small, hence supporting the validity of the approach. 
\section{Numerical results: single-electron spectra at RHIC and LHC}
For each explored case we generated an initial sample of $45\cdot 10^6$ $c\bar c$ and $b\bar b$ pairs, using the POWHEG code~\cite{POWHEG} with CTEQ6M PDFs. In the $AA$ case we introduced nuclear effects in the PDFs according to the EPS09 scheme~\cite{EPS09}; the quarks were then distributed in the transverse plane according to the nuclear overlap function ${dN/d\x_\perp\!\sim\!T_{AB}(x,y)}\!\equiv\!T_A(x\!+\!b/2,y)T_B(x\!-\!b/2,y)$.
At the proper-time $\tau\!\equiv{t^2\!-\!z^2}\!=\!\tau_0$ we started following the Langevin dynamics of the quarks until hadronization. The latter was modeled using Peterson fragmentation functions~\cite{peter}, with branching fractions into the different hadrons taken from Refs.~\cite{zeus,pdg}.
Finally each hadron was forced to decay into electrons with PYTHIA~\cite{Pythia}, using updated decay tables~\cite{pdg09}.
The $e$-spectra from $c$ and $b$ were then combined with a weight accounting for the respective total production cross-section and for the branching ratios of the various processes leading to electrons in the final state.
\begin{table}
\begin{center}
\begin{tabular}{|c|c|c|}
\hline
$\sqrt{s}_{NN}\!=\!200$ GeV & $\sigma_{c\bar c}$ ($\mu$b) & {$\sigma_{b\bar b}$ ($\mu$b)}\\
\hline
p-p & 254.14 & 1.769\\
\hline
Au-Au & 236.11 & 2.033\\
\hline
\hline
{$\sqrt{s}_{NN}\!=\!5.5$ TeV} & $\sigma_{c\bar c}$ (mb) & $\sigma_{b\bar b}$ (mb)\\
\hline
p-p & {3.0146} & 0.1872\\
\hline
Pb-Pb & {2.2877} & 0.1686\\
\hline
\end{tabular}
\begin{tabular}{|c|c|c|c|}
\hline
Hydro code & $\tau_0$ (fm/c) & $s_0$ (fm$^{-3}$) & $T_0$ (MeV)\\
\hline
ideal & 0.6 & 110 & 357\\
\hline
viscous & 1.0 & 83.8 & 333\\
\hline
\hline
Hydro code & $\tau_0$ (fm/c) & $s_0$ (fm$^{-3}$) & $T_0$ (MeV)\\
\hline
viscous & 0.1 & 1840 & 854\\
\hline
viscous & 1.0 & 184 & 420\\
\hline
\end{tabular}
\caption{Initialization for RHIC and LHC: the $c\bar c$ and $b\bar b$ production cross section given by POWHEG and the explored hydro scenarios.}\label{table:cross_hydro}
\end{center}
\end{table}
\begin{figure}[!h]
\begin{center}
\includegraphics[clip,width=0.495\textwidth]{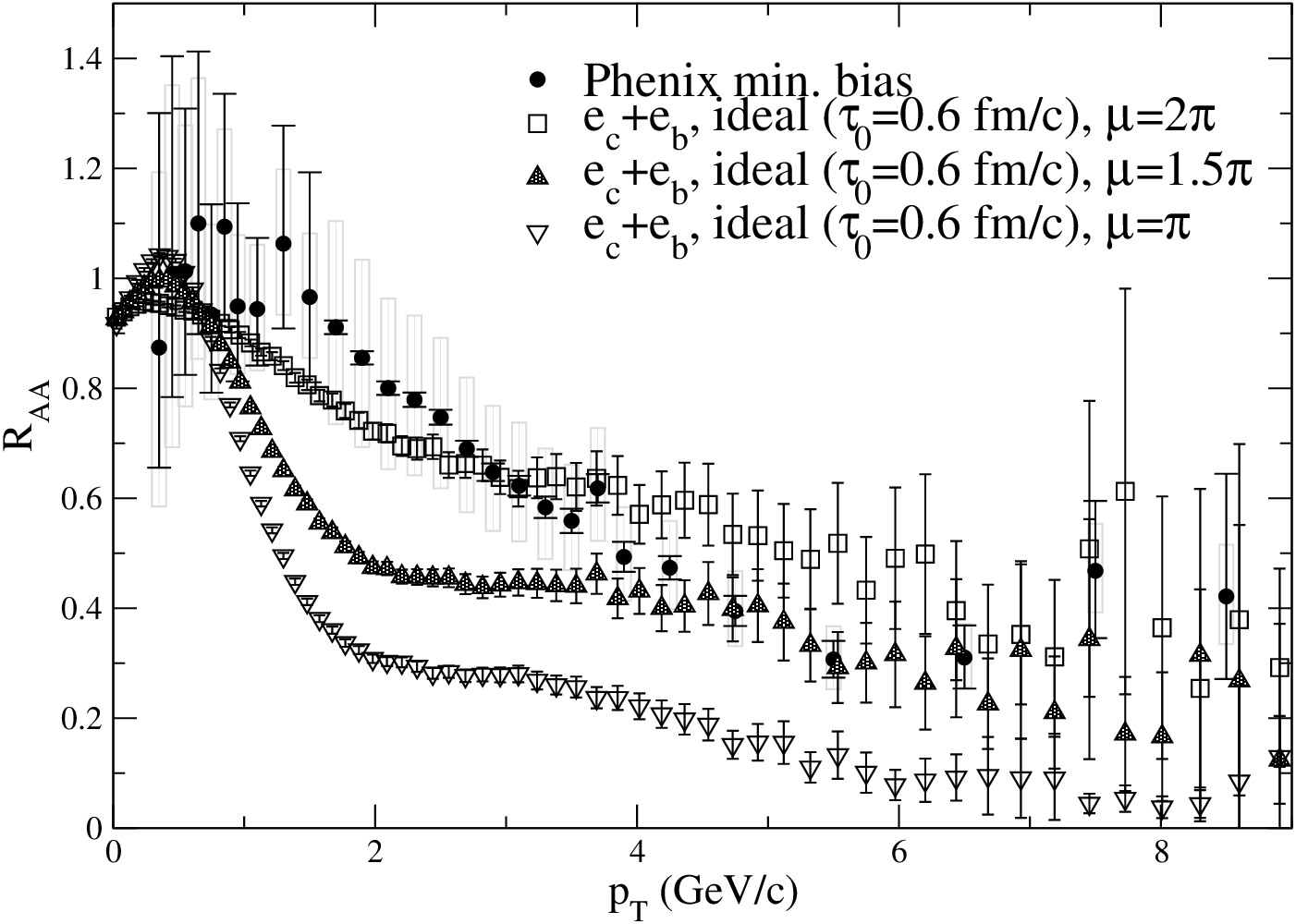}
\includegraphics[clip,width=0.495\textwidth]{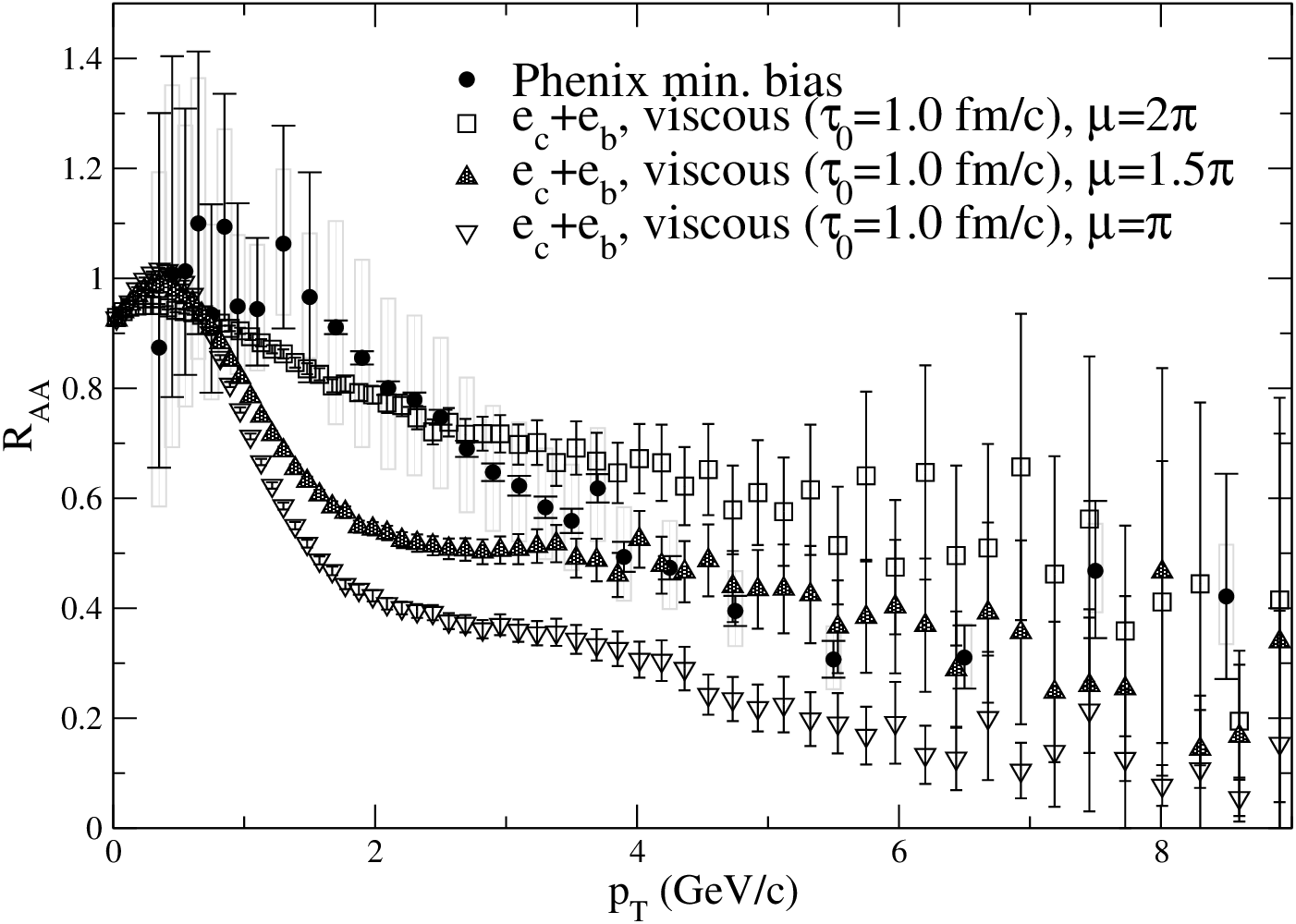}
\caption{The electron $R_{AA}$ for minimum bias Au-Au collisions (corresponding to an impact parameter $b\!=\!8.44$ fm) at $\sqrt{s}_{\rm NN}\!=\!200$ GeV compared with PHENIX results~\cite{Phenix}. The sensitivity to the scale $\mu\!=\!\pi T-2\pi T$ is displayed. The high-momentum region ($p_T\!\simge\!4$ GeV/c) is better reproduced with the intermediate coupling. Ideal (left panel) and viscous (right panel) hydro scenarios provide similar results.} 
\label{fig:RAA}
\end{center}
\end{figure}
\begin{figure}[!h]
\begin{center}
\includegraphics[clip,width=0.495\textwidth]{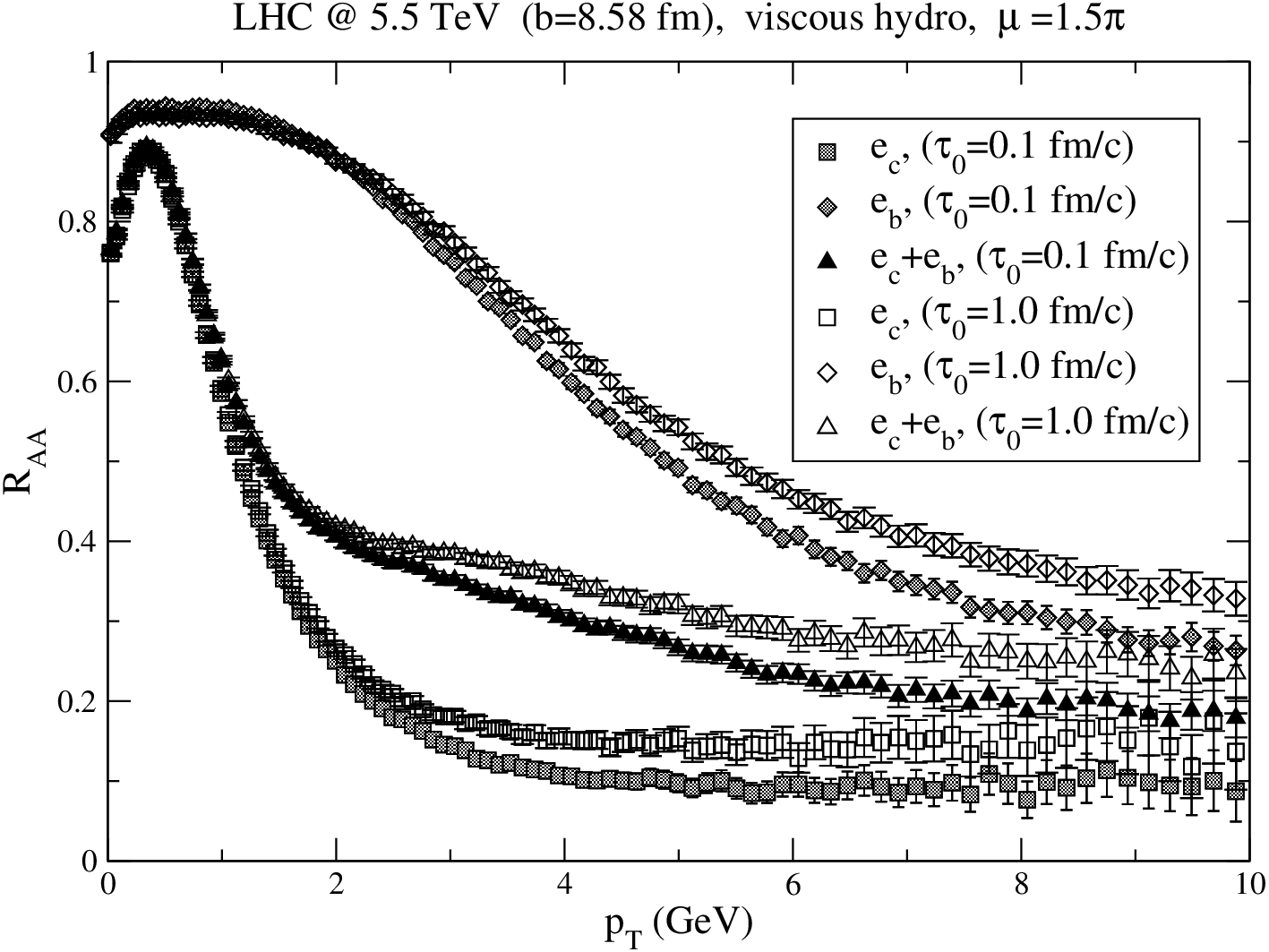}
\includegraphics[clip,width=0.495\textwidth]{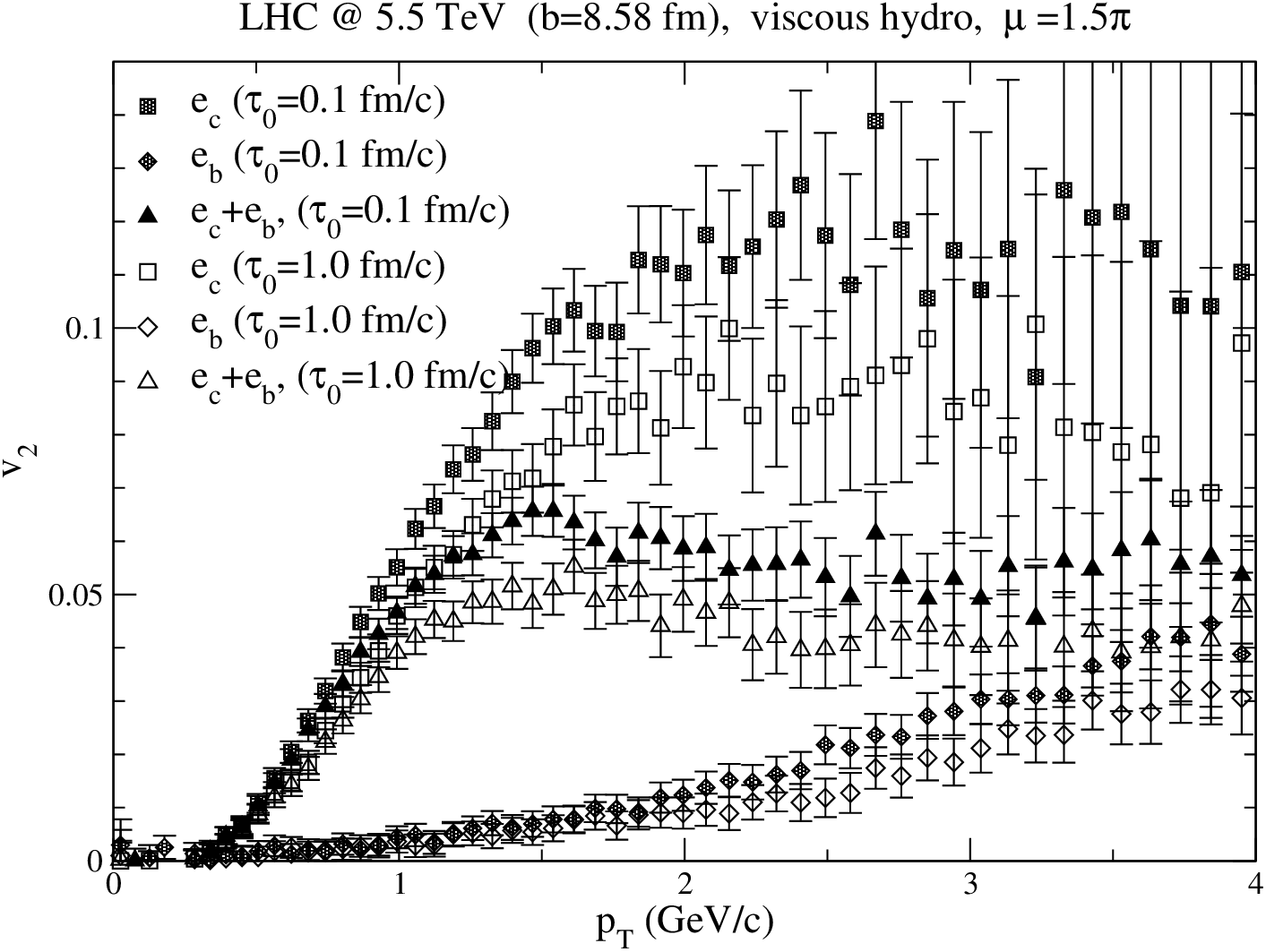}
\caption{The electron $R_{AA}$ and $v_2$ for minimum bias Pb-Pb collisions ($b\!=\!8.58$ fm) at LHC for the two hydro scenarios considered.}
\label{fig:LHC}
\end{center}
\end{figure}
We first address the RHIC case. The parameters characterizing the ``initial state'' are given in Table~\ref{table:cross_hydro}. In Fig.~\ref{fig:RAA} our results for the electron $R_{AA}$ in minimum-bias collisions are displayed and compared with PHENIX data~\cite{Phenix}. The dependence on the hydro scenario is quite small. The major uncertainty comes from the scale at which one evaluates the coupling $g$: for that we explore the range $\mu\!=\!\pi T-2\pi T$. The high-momentum region ($p_T\!\simge\! 4$ GeV/c) is better reproduced evaluating $\alpha_s$ at the scale $1.5\pi T$ ($\alpha_s\!=\!0.32$ at $T\!=\!300$ MeV). Hadronization via coalescence~\cite{rapp}, so far ignored, could improve the agreement at lower $p_T$. Results for $v_2$ can be found in Ref.~\cite{lange_pra}. 
%
\begin{figure}
\begin{center}
\includegraphics[clip,width=0.495\textwidth]{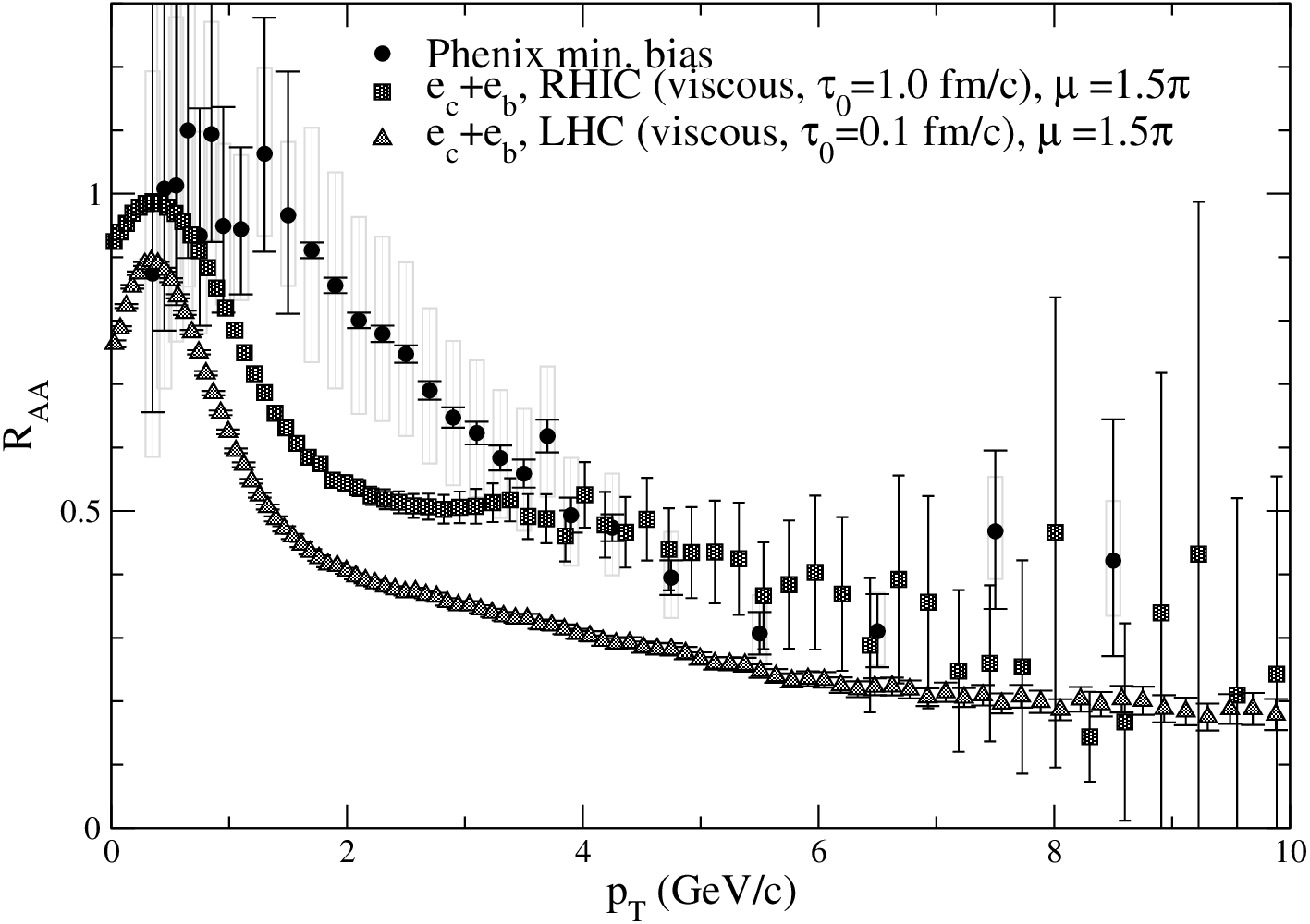}
\includegraphics[clip,width=0.495\textwidth]{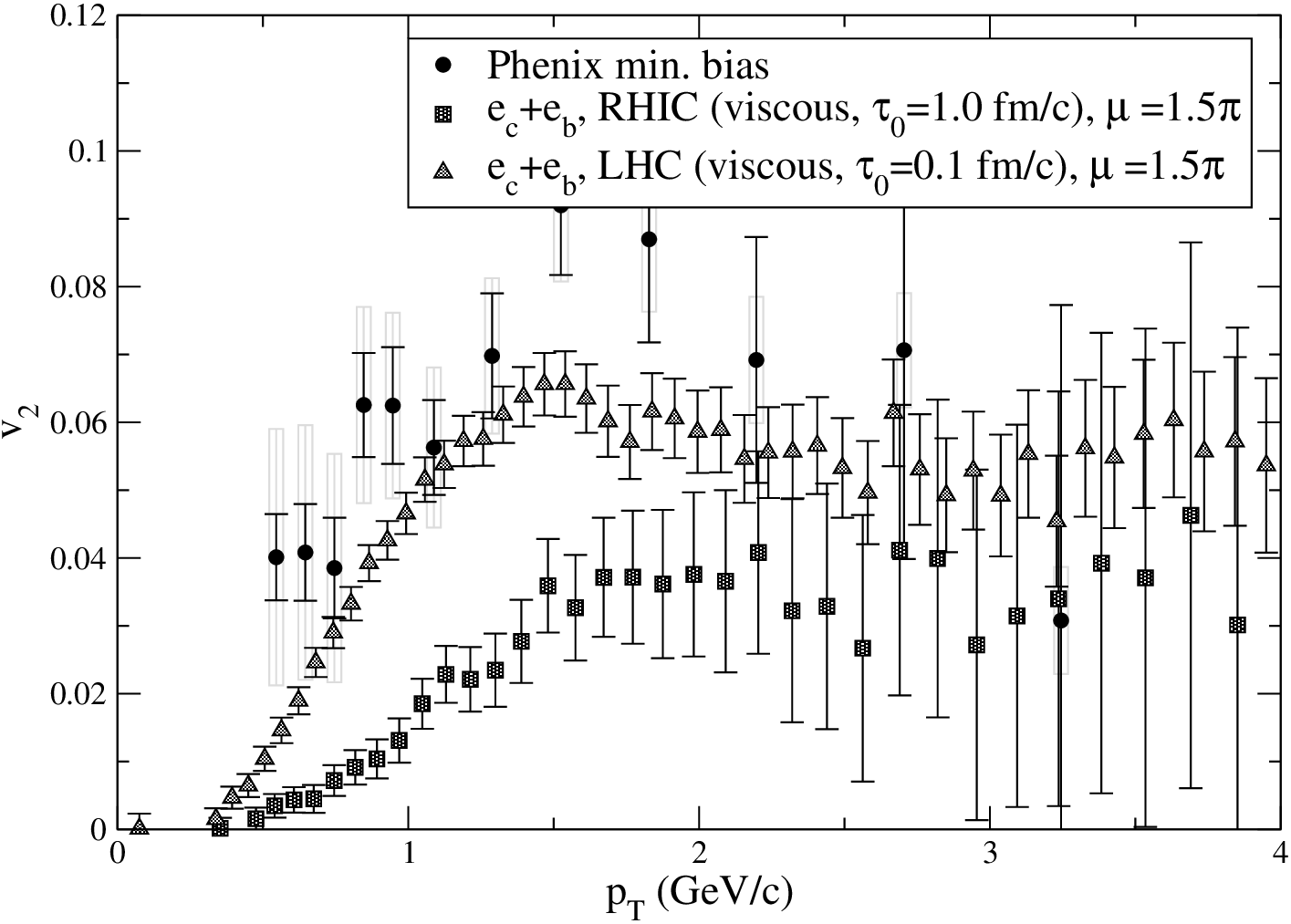}
\caption{The electron $R_{AA}$ and $v_2$ for minimum bias collisions at RHIC (Au-Au at $\sqrt{s}_{\rm NN}\!=\!200$ GeV) and LHC (Pb-Pb at $\sqrt{s}_{\rm NN}\!=\!5.5$ TeV).} 
\label{fig:RHICvsLHC}
\end{center}
\end{figure}
We address now the LHC case, described by the parameters in Table~\ref{table:cross_hydro}. Notice the dramatic effect of the nPDFs on the $c\overline{c}$ production. Our findings for $R_{AA}$ and $v_2$ are shown in Fig.~\ref{fig:LHC}. While the $b$-spectrum is simply suppressed at high-$p_T$, the charm -- beside a stronger quenching -- also displays a sizable flow. Finally in Fig.~\ref{fig:RHICvsLHC} we provide a comparison of our results for the inclusive ($c+b$) single-electron spectra for RHIC and LHC conditions. In the last case the larger initial temperature entails a stronger quenching and a more pronounced elliptic flow. The effect would be even stronger in the absence of initial state effects, responsible for a larger $\sigma_{b\bar b}/\sigma_{c\bar c}$ at LHC with respect to RHIC.
\bibliographystyle{elsarticle-num}

\begin{thebibliography}{00}
\bibitem{lange_hot} W.M. Alberico \emph{et al.}, arXiv:1007.4170 [hep-ph]. 
\bibitem{lange_pra} W.M. Alberico \emph{et al.}, arXiv:1009.2434 [hep-ph].
\bibitem{hira} Y. Akamatsu, T. Hatsuda, T. Hirano,
               Phys. Rev. C {\bf 79} 054907 (2009).
\bibitem{rapp} H. van Hees, V. Greco and R. Rapp,
               Phys. Rev. C {\bf 73}, 034913 (2006).
\bibitem{tea} G.D. Moore, D. Teaney,
              Phys. Rev. C {\bf 71}, 064904 (2005).
\bibitem{aic} P.B. Gossiaux and J. Aichelin,
              Phys. Rev. C {\bf 78}, 014904 (2008).
\bibitem{das} S.K. Das, Jan-e Alam, and P. Mohanty,
              Phys. Rev. C 80, 054916 (2009).ù
\bibitem{ito} K. Ito ``On stochastic differential equations''. Memoirs, American Mathematical Society 4, 1–51 (1951).
\bibitem{kolb1} P.F. Kolb, J. Sollfrank and U. Heinz,
                Phys. Rev. C {\bf 62}, 054909 (2000).
\bibitem{rom1} P. Romatschke and U.Romatschke,
               Phys. Rev. Lett. {\bf 99}, 172301 (2007). 
\bibitem{rom2} M. Luzum and P. Romatschke,
               Phys. Rev. C {\bf 78}, 034915 (2008). 
\bibitem{lange} A. Beraudo, A. De Pace, W.M. Alberico and A. Molinari,
                Nucl. Phys. A 831, 59 (2009).
\bibitem{pei} S. Peign\'e and  A. Peshier,
               Phys. Rev. D {\bf 77}, 114017 (2008).
\bibitem{com} B.L. Combridge,
              Nucl. Phys. B {\bf 151}, 429 (1979).
\bibitem{POWHEG} S. Frixione, P. Nason and G. Ridolfi,
                 JHEP 0709, 126 (2007).
\bibitem{EPS09} K.J. Eskola, H. Paukkunen and C.A. Salgado,
              JHEP 0904, 065  (2009).
\bibitem{peter} C. Peterson, D. Schlatter, I. Schmitt and P.M. Zerwas,
                Phys. Rev. D {\bf 27}, 105 (1983).
\bibitem{zeus} http://www-zeus.desy.de/zeus\_papers/ZEUS\_PAPERS/DESY-05-147.ps .
\bibitem{pdg} http://www.slac.stanford.edu/xorg/hfag/osc/PDG\_2009/\#FRAC .
\bibitem{Pythia} T. Sjostrand, S. Mrenna and P.Z. Skands,
                 JHEP 0605, 026 (2006).
\bibitem{pdg09} C. Amsler \emph{et al}. (PDG), Phys. Lett. B667, 1 (2008)
                and 2009 partial update.
\bibitem{Phenix} A. Adare \emph{et al.} (PHENIX Collaboration),
              Phys. Rev. Lett. {\bf 98},  172301 (2007).
\end{thebibliography}

\end{document}